# Effect of Channel Geometry and Flow Rates in Hydrodynamic Focusing on Impedance Detection of Circulating Tumor Cells

Hassan Raji[a], Iraj Dehghan Hamani[b]

[a]Department of Mechatronics Engineering, University of Tehran, Tehran, Iran
[b] Department of Mechanical Engineering, University of British Columbia, Canada

**Abstract**

Cells, other than their biological properties, have different electric and physical properties. In an impedance cytometer, cells should pass one by one in the detection region where pairs of electrodes are located. When cells are located between electrodes, the impedance changes, and this can be indicative of the presence of a cell. This is basically because the electric properties of cells are different from the medium between the electrodes which is important in determining the impedance. One of the most important aspects which influence the performance of an impedance cytometer performance is the microchannel design. In this work, in the first step, the microchannel was designed in a way to have the best detection in the impedance cytometer. In this regard, hydrodynamic focusing was selected to focus the population of cells entering from the inlet of the main channel. To find the optimal parameters of the microchannel, different geometry for the channel itself, along with flow rates and other parameters related to sheath flow were simulated. In the next step, impedance was measured in COMSOL for White blood cells, MCF7, and MDA-MB-231 breast cancer cells. The results show that by measuring the impedance of cells using the optimized channel design, CTCs can be successfully differentiated from WBCs.

## 1. Introduction

Microfluidic lab-on-a-chip technologies have been developed dramatically over last 20 years. The field started with analytical separations—mainly chip-based capillary electrophoresis as well as miniaturized chemical sensors accordingly, different platforms have been created to fulfilment operations such as sample take-up, sample preconditioning, reagent supply, metering, aliquoting, valving, routing, mixing, incubation, washing as well as analytical separations. There has been interest in miniaturized systems for medical diagnostics which often employ a microfluidic part [1]. In this regard, microfluidic cell counting plays a key role in medical diagnostics that can either utilize electrical detection [2] or optical detection [3]. These cell counters can be useful in diagnostics of various diseases such as Sickle cell Disease (SCD), Acute Myeloid Leukemia (AML), and metastatic cancers by detecting CTCs [4]. Hence, detection of CTCs is paramount to diagnose cancers in early stages, to evaluate cancer recurrence earlier and chemotherapeutic efficacy, and last but not the least the choice of anti-cancer drugs [5].

Microfluidic impedance cytometry is a simple, label-free, non-invasive technique which has experienced a significant interest over the past few years [6]. Alternatives that rely on optical microfluidics are generally more expensive and more complicated [7-8]. In impedance cytometry, pairs of microfabricated electrodes are located in a so-called detection region. Impedance of the medium between the electrodes is measured when there is no cell in the detection region. Once a cell flows between a pair of electrodes, the impedance changes. This is because of the difference between the dielectric properties such as size, permittivity, and conductivity of cells and the fluid that is flowing in the microchannel. A solution could be using channel with smaller dimensions but this may result in clogging and complexity in fabrication [9]. Cells should pass one by one in front of electrodes so that the impedance is owing to single cells rather than two or more cells. One way to focus cells in a way that only a single cells pass in a sequence is to use hydrodynamic focusing.

Ayliffe et al. proposed a leading example of microfluidic single-cell electric impedance spectroscopy [10]. This microfluidic system was made up of glass-polymer with integrated electrodes fabricated on the walls of the microchannel. Spectra over a frequency range up to 2 MHz was obtained for a variety of Phosphate Buffer Saline (PBS) solutions at different concentrations. Single human leukocytes and fish erythrocytes in a suspension were employed in this study. Nevertheless, low-frequency measurements of cells were dominated by the high capacitive response of their membrane. In addition, no information about the cell's internal properties was obtained. In another research, Gawad et al. developed a cytometer with a coplanar electrodes configuration for differential measurement of impedance [11]. This group showed the ability to perform simultaneous measurements at multiple frequencies up to 15 MHz and to differentiate beads from cells at a rate of 100 cell/s. The same group in another study replaced the coplanar electrodes with two pairs of opposing electrodes. These electrodes were located on both top and bottom of the channel. This microfluidic system was able to differentiate polystyrene beads and red blood cells based on the opacity, which is the ratio of impedance in two different frequencies. Holmes et al. detected white blood cells using a microfluidic impedance cytometer [12]. It was demonstrated that it was possible to identify and count T-lymphocytes, monocytes and neutrophils according to both membrane capacitance and size. Their data indicated that in the frequency range of up to 3 MHz, different types of cells are characterized by different impedance magnitudes. Desirable output data is dependent on having cells to flow in the middle of the main channel. Hydrodynamic focusing is an effective way for cell focusing in which a fluid named as sheath flow is used [2]. In this method, particles are typically focused to a stream traveling along the center of the main channel. This focusing is carried out by employing 2 side channels which make the cells to be located in the centerline of the main channel [13-14]. This method can be used to focus particles in micro-channels and by adjusting flow rate of side channels. Literature also describes different methods for fabricating micro/nano devices, such as soft lithography and additive manufacturing, as well as the potential of each [15-20].

On the other hand, multiphysics COMSOL simulations provide valuable information about behavior of the microsystem. Also, this data can be used while performing experiments because many aspects of the experiments is considered in this computational platform [21]. In this paper, a microfluidic system which uses hydrodynamic focusing and impedance cytometer is simulated. First, we propose a micro-channel which includes a main channel and two side channels with sheath fluid. In this part, we aim to have a better cell counter. In other words, it is tried to count cells one by one, to maximize the distance between cells while flowing in front of electrodes, and to minimize deviation of cells from centerline of the main channel. Then, pairs of electrodes are located where impedance of cells is measured in a range of frequency. The impedance of different WBCs and CTCs is obtained using the data in hydrodynamic focusing in COMSOL.

## 2. Microchannel design

It is necessary to pass the cells in front of the electrodes one at a time within a short time period in order to detect them. The main limitations of this method are the fabrication of narrow channels and the possibility of cells blocking the channel. Hence, to overcome the limitations and track impedance changes optimally, a hydrodynamic focusing method was used. Changes in impedance are mainly directed by three factors: cell positioning in the channel, cell diameter, and electrical properties of the individual cells. The flows are injected into the main flow via two side channels, resulting in particles being centralized within the primary flow. Figure (1) shows the preliminary micro-channel design as well as the parameters that affect cell centralization. A description of these parameters can be found in Table (1).

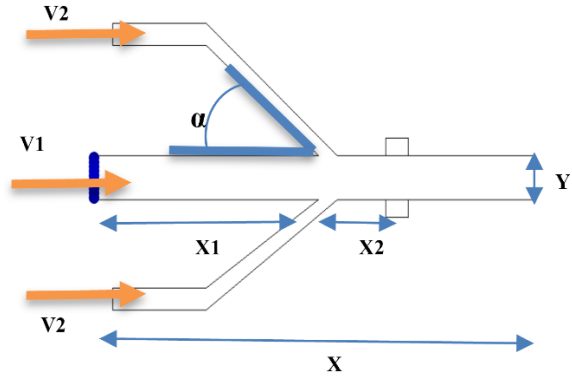

Fig. 1. primary design of micro-channel

Table 1. description of parameters

| Parameter | Description |
|---|---|
| Y | Channel width |
| X1 | Distance between basic flow inlet and sheath flow |
| X2 | Distance between sheath flow and electrodes |
| X | Channel length |
| α | Sheath flow angle |
| V1 | Basic flow velocity at the entrance |
| V2 | Sheath flow velocity at the entrance |

In our design, a laminar flow was considered in the micro-channels because of the low Reynolds number constraint. The fluid that drives the cells is governed by the characteristics of blood [22]. Also, the drag and Saffman forces are applied to particles in all simulations [23]. Three main types of white blood cells (WBC) including Lymphocyte, Monocyte and Neutrophil were considered in the simulations. The diameter, density and percentage of each of these three WBCs were extracted from [24] and are indicated in Table (2). This study simulates the MCF7 cell that is a kind of breast cancer cell along with other WBCs. MCF7 diameter is about 18µm and the conductivity and permittivity of this cell was considered 4 (S/m) and 50 respectively [25].

Table 2. Properties of white blood cells

| Cell type | Density(g/ml) | Diameter (µm) | Percentage |
|---|---|---|---|
| Lymphocyte | 1.073-1.077 | 6.58±0.7 | 33% |
| Monocyte | 1.067-1.077 | 9.26±0.72 | 5% |
| Neutrophil | 1.085-1.090 | 9.42±0.46 | 62% |

In order to optimize cancer cell detection, three parameters were considered. Table (3) indicates these three parameters and their definitions.

Table 3. Definition of optimization parameters

| Parameter | Definition |
|---|---|
| $\Delta y_{max}$ | The longest possible distance between particle and symmetry axis of the main channel |
| $\Delta x_{min}$ | The shortest possible distance between two particles while going in front of electrodes |
| T | Sensing time |

The channel design is based on three cost functions: decreasing $\Delta y_{max}$, for having a better centralization of particles; increasing $\Delta x_{min}$, in order to prevent particles overlapping and eliminating the particles presence effect on each other during the sensing time; and reducing T, for having a fast detection performance. Thus, the microchip geometry is optimized based on these three parameters.

As indicated in [26], the best angle for side channels is 30 degrees to have the best cell centralization in the main channel. Furthermore, by decreasing the channel width (Y), the centralization would be improved; however, this could cause particle blockage. Therefore, a 150µm channel width (Y=150µm) is considered. The channel length should be longer for increasing $\Delta x_{min}$, but a 1cm (X=1cm) channel length was considered due to have a smaller device. Moreover, the electrodes in this design are located close to the end of the channel.

Three positions are considered for optimizing the location of side channels sheath flow with respect to the main flow (Figure (2)). Table (4) shows the simulations' results in which V1=100µm/s, V2=500µm/s and Y=150µm are applied. It is seen that by increasing the distance between the sheath flow location and the main flow entrance, more sensing time is required. On the other hand, it improves cell centralization, so the X1=5mm is chosen as the optimum place for sheath flow. Then, for optimizing the inlet velocity of main flow (V1), the inlet velocity of side channels are considered 1000µm/s, and the optimization parameters are calculated while X1=5mm and X=1cm and the V1 is changed from 100 to 750 µm/s. Table (5) shows the computed values of the three different optimization parameters to find the best V1 in our design.

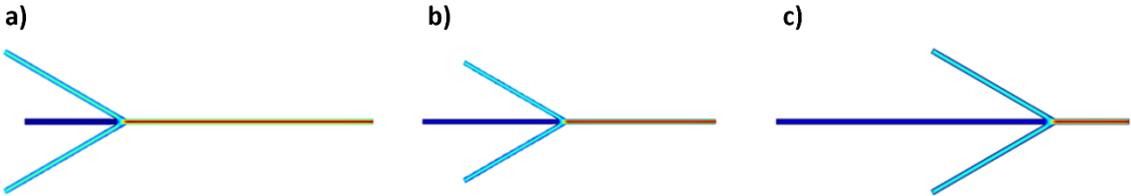

Fig. 2. Position of the side channels with respect to the main channel a) X1=3mm, b) X1=5mm, c) X1=7mm

Table 4. Simulation results regarding the location of sheath flow

| Sheath flow location (X1) | $\Delta y_{max}$ | $\Delta x_{min}$ | T |
|---|---|---|---|
| 3mm | 4.65 | 1.2 | 40 |
| 5mm | 4.86 | 8 | 70 |
| 8mm | 5.2 | 18.2 | 110 |

Table 5. Simulation results for inlet velocity of main flow

| Inlet velocity of main flow V1 (µm/s) | $\Delta y_{max}$ | $\Delta x_{min}$ | T |
|---|---|---|---|
| 100 | 2.6 | 1.3 | 65 |
| 250 | 40.8 | 5.2 | 35 |
| 500 | 10 | 31.2 | 25 |
| 750 | 13.1 | 51 | 15 |

We considered several different cases for optimizing channel width and flow velocity, and we selected the best one resulting in better particle centralization and sensing time. By increasing V1, the sensing time decreases when the $\Delta x_{min}$ and $\Delta y_{min}$ are increased; and since the smaller $\Delta y_{max}$ is desirable, V1=500 µm/s is selected. It is notable to mention that the cells travel in the channel based on the velocity profile, so some particles near the channel wall travel slowly and arrive later than other particles, which helps to have a better $\Delta x_{min}$. Figure (3) indicates cells travelling in the channel based on the flow velocity profile.

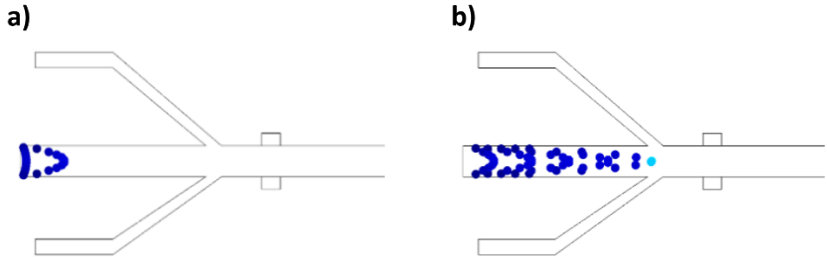

Fig. 3. cells travelling in the channel (a) from one side of main channel and (b) reach to the side channel over time

For optimizing the sheath flow velocity at the entrance (V2), the V1=500 µm/s, X1=5m are considered to be constant and V2 is changed from 1000 to 5000 µm/s. Table (6) shows the results of optimization parameters. By increasing V2, T and $\Delta y_{max}$ decrease while $\Delta x_{min}$ increases, which is in line with our optimization goals. Hence, V2=3000 (µm/s) is selected.

Table 6. Simulation results for inlet velocity of side flow

| Inlet velocity of side flow V2 (µm/s) | $\Delta y_{max}$ | $\Delta x_{min}$ | T |
|---|---|---|---|
| 1000 | 10.1 | 31 | 25 |
| 1250 | 8.8 | 283 | 20 |
| 1500 | 7.1 | 331 | 17 |
| 2000 | 5.6 | 362 | 15 |
| 3000 | 4.3 | 502 | 14 |
| 5000 | 2.1 | 800 | 13 |

In addition to finding optimum parameters for V1, V2, and X1, three different channel widths are explored. V1=500 µm/s and V2=5000 µm/s are set as the velocity in the main and side channels. Table (7) shows that by decreasing the width, the shortest distance between two particles ($\Delta x_{min}$) decreases, but the centralization process improves (smaller $\Delta y_{max}$). Also, a smaller channel width makes a better electrical detection. Thus, Y=50 µm is considered for the final simulations.

Table 7. simulation results for different channel widths

| Channel width Y (µm) | $\Delta y_{max}$ | $\Delta x_{min}$ | T |
|---|---|---|---|
| 150 | 2.2 | 805 | 13 |
| 100 | 1.7 | 260 | 13 |
| 50 | 1.4 | 96 | 13 |

It is essential that the microchip design considers the following geometry and parameters in order to achieve the best results in cell detections (having an acceptable centralization and sensing time): V1=500µm/s, V2=5000µm/s, X1=5mm, and Y=50µm.

## 3. Electrical detection of cancer cells

COMSOL simulations are performed on the cross section of a channel with 500 um length, 50 um width using opposing electrodes. The electrodes embedded onto the top and bottom surfaces of the microfluidic channel are each 15 um wide. Impedance between the electrodes was measured in two scenarios: when no cell was between the electrodes (See Figure (4)) and when MCF-7 and WBCs are between the electrodes (See Figure (5)). In the channel medium, the conductivity was set to the blood fluid conductivity, while at the channel wall boundary, the insulator was selected. Figures 4 and 5 illustrate the potential gradient across the channel and variation of the field intensity in the channel due to the dielectric properties of the MCF-7.

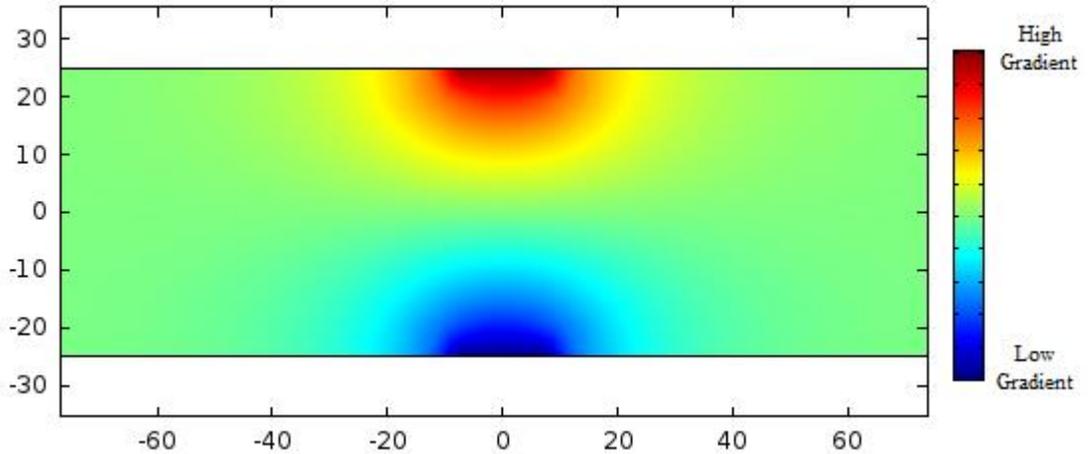

Fig. 4. Gradient of potential in the channel when there is no cell between the electrodes

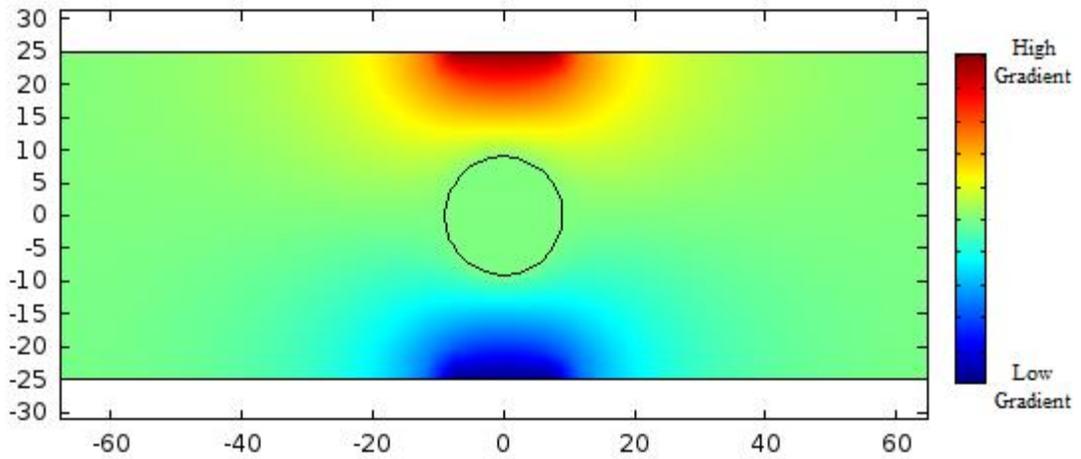

Fig. 5. Change in potential intensity in the channel due to the dielectric properties of MCF7

The passage of a cell between electrodes alters the electrical impedance measured between the electrodes. The electrodes of our biosensor are tested with WBCs and MCF-7 placed between them. By applying a voltage between the electrodes and sweeping the frequency, the impedance of WBCs and MCF7 is obtained as it is seen in figure (6). A significant difference between the cancer cell and other blood cells shows the device's ability to detect this type of cancer cell over a wide range of frequencies. Even for the same type of blood cells, impedances differ, but by using mean values, the impedance of MCF-7 can be measured properly as well.

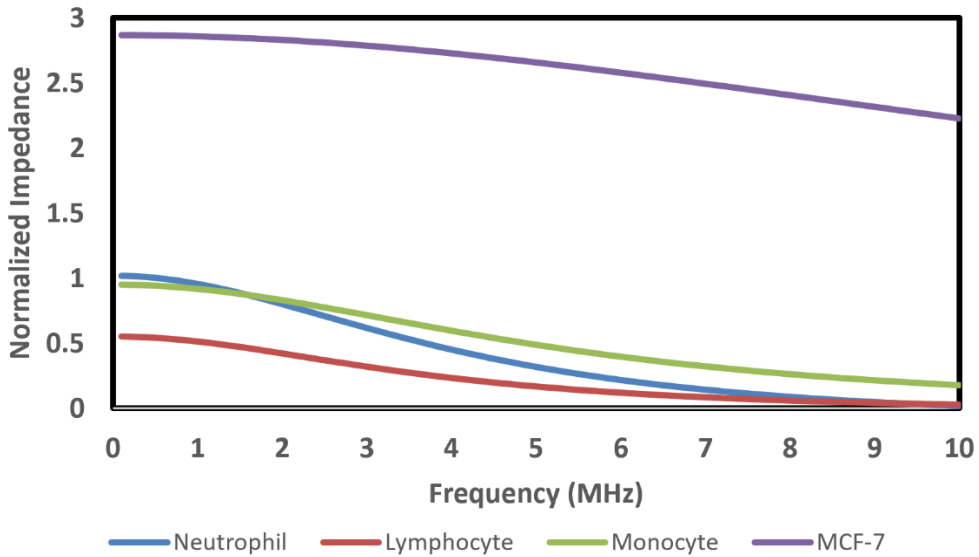

Fig. 6. Normalized impedance of WBCs and MCF-7 between 0 to 10 MHz

Normalized impedance is proportional to the difference of the amplitude of impedance when a single cell is between electrodes and the amplitude of impedance when no cell is located between electrodes. The normalized

88

impedance for WBCs and MDA-MB-231 is shown in Figure (7) using the same simulation properties yet considering relevant properties of MDA-MB-231 in COMSOL to evaluate the performance of the device also for this type of CTC. MCF-7 has a higher impedance than MDA-MB-231, but still has a different value from WBCs in various frequencies that makes it suitable for use as a biosensor.

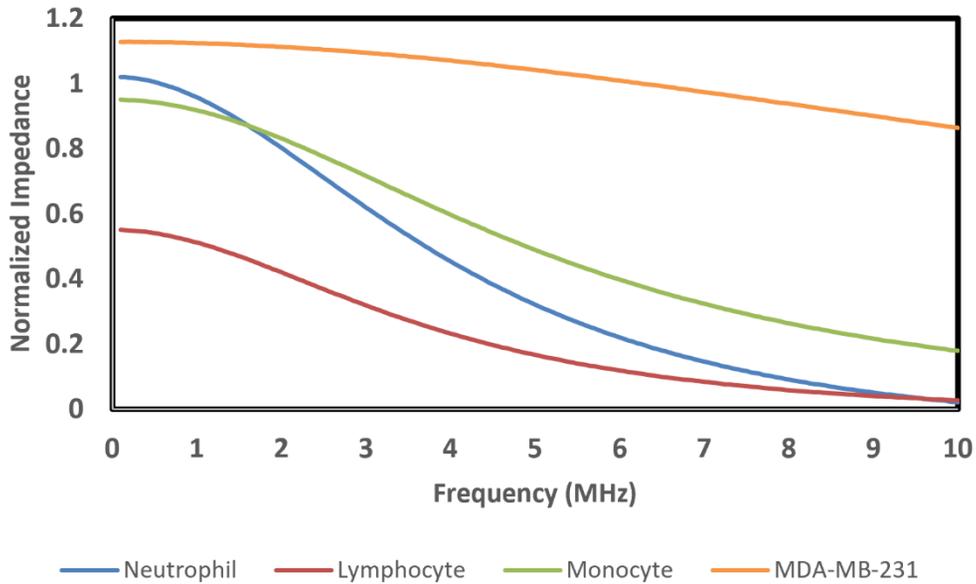

Fig. 7. Normalized impedance of white blood cells and MDA-MB-231

## 4. Conclusion

Using impedance cytometry, we demonstrated how subpopulations of circulating tumor cells can be identified from a mix of WBCs. Specifically, MCF-7 and MDA-MB-231 cancer cells were detected in blood. A pair of opposing electrodes were used to determine the difference in impedance between WBCs and CTCs. It represents a simple and effective way of detecting CTCs in mixed populations of cells. The impedance of MCF-7 is more than 3x that of all WBC types in a frequency range up to 10 MHz. The study also examines how channel parameters, such as channel geometry or flow velocity, can affect the performance of an impedance biosensor. Three major characteristics of a microfluidic impedance cytometer were selected as measures to optimize channel parameters. We optimized the flow velocity in the main and side channels, channel width, and relative positioning of side channels in the hydrodynamic focusing. This paper shows how these parameters change detection time, deviation of cells in the main channel and distance between cells.